\documentclass[twocolumn]{aastex7}
\usepackage{amsmath}

\usepackage{CJK}

\newcommand{\beq}{\begin{equation}}
\newcommand{\eeq}{\end{equation}}
\usepackage{orcidlink}
\newcommand{\orcid}[1]{\href{https://orcid.org/#1}{\orcidlink{#1}}}

\begin{document}
\begin{CJK*}{UTF8}{gbsn}
\title{Testing the Shock-cooling Emission Model from Star-Disk Collisions for Quasiperiodic Eruptions}

\author[orcid=0009-0000-5741-4456]{Wenyuan Guo (郭文渊) \orcid{0009-0000-5741-4456}}
\affiliation{School of Physics and Astronomy, Sun Yat-sen University, Zhuhai 519082, China}
\affiliation{CSST Science Center for the Guangdong-Hong Kong-Macau Greater Bay Area, Sun Yat-sen University, Zhuhai 519082, China}
\email[show]{guowy25@mail2.sysu.edu.cn}  

\author[orcid=0000-0001-5012-2362]{Rong-Feng Shen (\text{申荣锋}) \orcid{0000-0001-5012-2362}}
\affiliation{School of Physics and Astronomy, Sun Yat-sen University, Zhuhai 519082, China}
\affiliation{CSST Science Center for the Guangdong-Hong Kong-Macau Greater Bay Area, Sun Yat-sen University, Zhuhai 519082, China}
\email[show]{shenrf3@mail.sysu.edu.cn} 


\begin{abstract}

Quasiperiodic eruptions (QPEs), the repeated outbursts observed in soft X-ray bands, have attracted broad interest, but their physical origin is under debate. One of the popular models, the star-disk collision model, suggests that QPEs can be produced through periodic collisions of an orbiting star with the accretion disk of a central black hole (BH). However, previous tests of the star-disk collision model mainly focus on the timing analysis. Other observed properties, such as peak luminosities $L_{\rm{p}}$, durations $t_{\rm{e}}$, and radiation temperatures $T_{\rm{p}}$ of the eruptions, are not systematically investigated. For a sample of six QPE sources and two QPE-like sources, we test the shock-cooling emission model from star-disk collisions by using these observables to derive the constraints on the stellar radius $R_\star$. We find that, except for two sources (eRO-QPE3 and eRO-QPE4), the rest of the sample either has no allowed $R_\star$ to simultaneously reproduce the observed $L_{\rm{p}}$ and $t_{\rm{e}}$ or the required $R_\star$ is too large to avoid being disrupted by the central BH. For the two exceptions, a stellar radius of the order of $1\ R_{\rm{\odot}}$ is necessary to satisfy all the constraints. Another issue with the simplest version of this model is that it predicts $k T_{\rm{p}} \sim 10\ \rm{eV}$, one order of magnitude lower than the observed value.

\end{abstract} 


\section{Introduction}

The probe of massive black holes (BHs) is one of the most attractive subjects in astrophysics. This helps answer whether there is at least a supermassive BH (SMBH) located in each galaxy's center. Past efforts to investigate the population of SMBHs have been based on observing the orbital motion of surrounding stars or some BH-accretion-related high-energy phenomena. In the near future, gravitational wave detectors have the potential to detect SMBHs in binary systems.

Recently, a new class of repeating soft X-ray outbursts called quasiperiodic eruptions (QPEs) has been discovered. Motivated by the first reported QPE, GSN 069 \citep{miniutti19, miniutti23b, miniutti23a}, RX J1301 and eRO-QPE1-4 are discovered \citep{sun17, Giustini20, Arcodia21, Arcodia24a, Arcodia24b, joheen24}. They exhibit luminous flares with the peak luminosity of $L_{\rm{p}} \approx 10^{42} \ \rm{erg\,s^{\rm{-1}}}$, the duration of $t_{\rm{e}} \approx 1\ \rm{hr}$, and recurring over a period of $P \approx 10\ \rm{hr}$. Based on the fitting of the spectrum, their peak temperature $k T_{\rm{p}} \approx 100\ \rm{eV}$. They are located at the nuclei of their host galaxies with a quiescent luminosity of $L_{\rm{Q}} \approx \rm{10^{41}\ \rm{erg\,s^{\rm{-1}}}}$, thus potentially linked to a massive BH with the mass of $M_{\rm{h}} \approx 10^{6} M_{\rm{\odot}}$. Also, due to periodic behaviors, QPE-like events like ASASSIN14-ko and Swift J0230 \citep[hereafter called J0230 for short;][]{payne21, payne22, payne23, huang23, linial24, evans23, guolo24} are expected to originate from the same physical process as QPEs.

On the basis of observations, several theoretical models have been proposed to interpret QPEs. Generally, they can be classified into four kinds: (1) disk instability model \citep{raj21, pan22, kaur23, sniegowska23}, (2) self-lensing model in massive BH binary \citep{ingram21}, (3) star-disk collision model \citep{dai10, pihajoki16, sukova21, xian21, franchini23, lu23, linial23, linial24, tagawa23, linial25}, (4) mass transfer model between a stellar companion and the SMBH, such as via Roche lobe overflow \citep{metzger22, linial23a}, repeated tidal stripping \citep{king20, king22, king23, zhao22, chen23, wu24}, and stream-stream collision induced by mass transfer \citep{krolik22}. Here we focus on the star-disk collision model.

For a star orbiting around the SMBH with an accretion disk, the star will keep colliding with a disk at the crossing points between the misaligned orbit and the disk plane. The flare comes out of the ejecta shocked by the collision. \cite{dai10} first investigated the star-disk collision model and built a connection between the quasiperiodic behavior of flares and the star's orbital parameters. After the discovery of GSN 069 \citep{miniutti19}, the star-disk collision model was applied to interpret the QPEs. Several works have been dedicated to the timing analysis of QPE \citep{xian21,franchini23,linial23,joheen24,zhou24a,zhou24b,zhou24c, miniutti25}, which aims to reproduce the high-low peak luminosity pattern and the long-short pattern of QPE by various precessions.

Besides timing, the observed radiative property is also important for testing the model. By estimating $L_{\rm{p}}$, $T_{\rm{p}}$, $t_{\rm{e}}$ and other observation properties, \cite{linial23} analytically studied the X-ray emission of the star-disk collision model. 
\cite{tagawa23} investigate the breakout emission in the star-disk collision model, incorporating the effect of orbital inclination. \cite{vurm24} explored the photon production process during the star-disk collisions and the radiative properties of the shocked material when it freely expands from the disk surface.

As an improved version of the star-disk collision model, the stream-disk collision scenario was studied by \cite{lu23}, \cite{linial24} and \cite{linial25}. Here, the debris stream was shed from the star either due to heating-induced inflation and ablation resulting from a previous collision, as shown in the hydrodynamic simulation by \cite{yao25}, or because of the tidal stripping by the SMBH \citep{huang23}.


However, the star-disk collision model is still not fully constrained under the observation data, much less the validation of its improved version: the stream-disk collision scenario. For example, it is difficult to find a satisfactory quiescent-level mass accretion rate to reproduce the observed $L_{\rm{p}}$, $T_{\rm{p}}$ and $t_{\rm{e}}$ simultaneously, as was shown in \cite{linial23}.


In this paper, we carefully check the estimations in \cite{linial23} and try to obtain a more holistic constraint on the shock-cooling emission model from star-disk collisions based on various observations of QPEs and QPE-like events. \S \ref{sec:2} presents a sample of QPE and QPE-like events to be used for the test. The theoretical framework of the model and the test method we apply are presented in \S \ref{sec:3}, and the results are given in \S \ref{sec:4}. Some additional effects and possible scenarios are discussed in $\S$ \ref{sec:5}. Finally, we draw our conclusions in \S \ref{sec:6}.

\section{The Sample of QPE and QPE-like Events}
\label{sec:2}

\begin{table*}[htbp!]
\centering
\renewcommand{\arraystretch}{1.3}
\begin{tabular}{l|c|c|c|c|c|c|c|c|c}
\hline
\hline
$\ $&P&\multicolumn{2}{c|}{$L_{\rm{p}}\ \left(10^{42}\ \rm{erg\,s^{\rm{-1}}} \right)$}&\multicolumn{2}{c|}{$t_{\rm{e}}$}&\multicolumn{2}{c|}{$L_{\rm{Q}}\ \left(10^{41}\ \rm{erg\,s^{\rm{-1}}}\right)$}&$kT_{\rm{p}}$&$M_{6}$\\[2pt]

\cline{3-9}
$\rm{Source}$&\fontsize{6}{14}$\ $&\fontsize{6}{14}$\rm{Range}$&\fontsize{6}{14}$\rm{Typical\ Value}$&\fontsize{6}{14}$\rm{Range}$&\fontsize{6}{14}$\rm{Typical\ Value}$&\fontsize{6}{14}$\rm{Range}$&\fontsize{6}{14}$\rm{Typical\ Value}$&\fontsize{6}{14}$\left(\rm{eV}\right)$&\fontsize{6}{14}$\ $\\
\hline
\fontsize{8}{14}$\rm{GSN}\ \rm{069}\ {\left(a\right)}$&\fontsize{8}{14}$8.6\ \rm{hr}$&\fontsize{8}{14}$1$-$5$&\fontsize{8}{14}$3$&\fontsize{8}{14}$(0.7$-$1.2)\ \rm{hr}$&\fontsize{8}{14}$0.94\ \rm{hr}$&\fontsize{8}{14}$200$-$300$&\fontsize{8}{14}$250$&\fontsize{8}{14}$70$-$120$&\fontsize{8}{14}$0.3$-$4$\\
\hline
\fontsize{8}{14}$\rm{RX}\ \rm{J1301}\ {\left(b\right)}$&\fontsize{8}{14}$4.7\ \rm{hr}$&\fontsize{8}{14}$1$-$3$&\fontsize{8}{14}$1.0$&\fontsize{8}{14}$(0.3$-$0.5)\ \rm{hr}$&\fontsize{8}{14}$0.4\ \rm{hr}$&\fontsize{8}{14}$0.3$-$2$&\fontsize{8}{14}$0.6$&\fontsize{8}{14}$94$-$330$&\fontsize{8}{14}$0.8$-$2.8$\\
\hline
\fontsize{8}{14}$\rm{eRO}$-$\rm{QPE1}\ {\left(c\right)}$&\fontsize{8}{14}$18 \ \rm{hr}$&\fontsize{8}{14}$0.37$-$43$&\fontsize{8}{14}$4.0$&\fontsize{8}{14}$(6.6$-$8.6)\ \rm{hr}$&\fontsize{8}{14}$7.5\ \rm{hr}$&\fontsize{8}{14}$0.10$-$0.66 $&\fontsize{8}{14}$0.38$&\fontsize{8}{14}$85$-$200$&\fontsize{8}{14}$ 6 \times 10^{-5}$-$4$\\
\hline
\fontsize{8}{14}$\rm{eRO}$-$\rm{QPE2}
\ {\left(d\right)}$&\fontsize{8}{14}$2.4\ \rm{hr}$&\fontsize{8}{14}$0.1$-$1.2$&\fontsize{8}{14}$0.35$&\fontsize{8}{14}$(0.4$-$0.5)\ \rm{hr}$&\fontsize{8}{14}$0.45\ \rm{hr}$&$-$&\fontsize{8}{14}$0.4$&\fontsize{8}{14}$150$-$240$&\fontsize{8}{14}$3 \times 10^{-5}$-$3$\\
\hline
\fontsize{8}{14}$\rm{eRO}$-$\rm{QPE3}\ {\left(e\right)}$&\fontsize{8}{14}$20\ \rm{hr}$&\fontsize{8}{14}$0.4$-$1$&\fontsize{8}{14}$0.6$&\fontsize{8}{14}$(2$-$2.7)\ \rm{hr}$&\fontsize{8}{14}$2.3\ \rm{hr}$&\fontsize{8}{14}$0.2$-$30$&\fontsize{8}{14}$2.5$&\fontsize{8}{14}$70$-$140$&\fontsize{9}{14}$0.46$-$40$\\
\hline
\fontsize{8}{14}$\rm{eRO}$-$\rm{QPE4}\ {\left(f\right)}$&\fontsize{8}{14}$14\ \rm{hr}$&\fontsize{8}{14}$4$-$15$&\fontsize{8}{14}$7.9$&\fontsize{8}{14}$(0.9$-$1.1)\ \rm{hr}$&\fontsize{8}{14}$1\ \rm{hr}$&\fontsize{8}{14}$1$-$100$&\fontsize{8}{14}$10$&\fontsize{8}{14}$79$-$127$&\fontsize{8}{14}$13$-$120$\\
\hline
\fontsize{8}{14}$\rm{ASASSN}$-$\rm{14ko}\ {\left(g\right)}$&\fontsize{8}{14}$115\ \rm{days}$&\fontsize{8}{14}$12$-$200$&\fontsize{8}{14}$50$&\fontsize{8}{14}$(10$-$36)\ \rm{days}$&\fontsize{8}{14}$23\ \rm{days}$&\fontsize{8}{14}$20$-$200$&\fontsize{8}{14}$63$&\fontsize{8}{14}$0.3$-$2.5$&\fontsize{8}{14}$12$-$159$\\
\hline
\fontsize{8}{14}$\rm{J0230} \ {\left(h\right)}$&\fontsize{8}{14}$22\ \rm{days}$&\fontsize{8}{14}$0.7$-$6$&\fontsize{8}{14}$2$&\fontsize{8}{14}$(2.4$-$6.5)\ \rm{days}$&\fontsize{8}{14}$4 \ \rm{days}$&\fontsize{8}{14}$1$-$2$&\fontsize{8}{14}$1.5$&\fontsize{8}{14}$150$-$200$&\fontsize{8}{14}$1.5$-$10$\\
\hline
\hline
\end{tabular}

\caption{The observed quantities regarding the radiation properties ($P$, $t_{\rm{e}}$, $L_{\rm{p}}$, $L_{\rm{Q}}$ and $kT_{\rm{p}}$) and the BH masses $M_{\rm{h}}$ for a sample of QPEs and QPE-like events. The symbols in the Source column are as follows: (a) \cite{miniutti19, miniutti23a, miniutti23b}, (b) \cite{sun17}, \cite{Giustini20}, (c) \cite{Arcodia21}, \cite{joheen24}, (d) \cite{Arcodia21}, \cite{pasham24a}, \cite{Arcodia24a}, (e) \cite{Arcodia24b}, (f) \cite{Arcodia24b}, (g) \cite{payne21, payne22, payne23}, \cite{huang23}, \cite{linial24}, (h) \cite{evans23}, \cite{guolo24}.}
\label{table:1}
\end{table*}

QPE and QPE-like events are outstanding for their luminous flares with repeated pulse-like light-curve profiles. In this section, we summarize these observed properties of QPE and QPE-like events, which are used in the constraints of $r_\star$.

GSN 069 shows a sequence of soft X-ray eruptions lasting for $t_{\rm{e}} \approx 1.1\  \rm{hr}$ with $P \approx 8.5\ \rm{hr}$ and $L_{\rm{p}} \approx \left(10^{42} - 10^{43}\right)\ \rm{erg\,s^{\rm{-1}}}$ \citep{miniutti19, miniutti23a, miniutti23b}. Between eruptions, $L_{\rm{Q}} = \left(2-3\right) \times 10^{43}\ \rm{erg\,s^{\rm{-1}}}$. Spectral fitting gives $k T_{\rm{p}} \approx \left(70-120\right)\ \rm{eV}$. Based on the DISKBB model fitting, \cite{miniutti19, miniutti23a,miniutti23b} inferred $M_{\rm{h}} \approx \left( 2-4 \right) \times 10^6 \rm{M}_{\rm{\odot}}$.

RX J1301 has shown three rapid flares in $2019$ and one flare in $2000$. These flares last for $t_{\rm{e}} = \left(1-2\right)\ \rm{ks}$ with $L_{\rm{p}} \approx 10^{42}\ \rm{erg\,s^{\rm{-1}}}$ and $P = 17\ \rm{ks}$ \citep{Giustini20}. During the quiescent state, $L_{\rm{Q}} = \left(0.3-2\right) \times 10^{41}\ \rm{erg\,s^{\rm{-1}}}$ \citep{sun17, Giustini20}. The mass of BH inferred from the UV/X-ray analysis is $M_{\rm{h}} = \left(8-28\right) \times 10^{5}\ M_{\rm{\odot}}$ \citep{sun17}. 

eRO-QPE1 is first reported for its luminous flares lasting for $t_{\rm{e}} = \left(6.6-8.6\right)\ \rm{hr}$ with $P = 18 \ \rm{hr}$ \citep{Arcodia21} and the later $3.5\ \rm{yr}$ monitoring presented by \cite{joheen24} shows its large variation in light curve. Generally, its various flares peak with $L_{\rm{p}} = \left(8-200\right) \times 10^{41} \ \rm{erg\,s^{\rm{-1}}}$ and $L_{\rm{Q}} = \left(1-6.6\right) \times 10^{40}\ \rm{erg\,s^{\rm{-1}}}$. eRO-QPE2 is reported by \cite{Arcodia21, Arcodia24a} and \cite{pasham24a} as a luminous QPE source with $P = 2.4\ \rm{hr}$. It rises to a high state for $t_{\rm{e}} = \left(24-30\right)\ \rm{min}$ with $L_{\rm{p}} = \left(0.1-1.2\right)\times10^{42}\ \rm{erg\,s^{\rm{-1}}}$. In the quiescent state, $L_{\rm{Q}} = 4\times10^{40}\ \rm{erg\,s^{\rm{-1}}}$. The central BH masses were estimated for these two sources based on disk luminosity, duration, and period with adopting different viscosity parameters $\alpha$.

eRO-QPE3 and eRO-QPE4 are reported by \cite{Arcodia24b}. Based on observations, eRO-QPE3 peaks with $L_{\rm{p}} = \left(0.4-1\right) \times 10^{42}\ \rm{erg\,s^{\rm{-1}}}$ with $P = 20\ \rm{hr}$ and $L_{\rm{Q}} = 1.6 \times 10^{40}\ \rm{erg\,s^{\rm{-1}}}$. For eRO-QPE4, the eruptions repeated every $14\ \rm{hr}$ with $L_{\rm{p}} = \left(4-15\right)\times 10^{42} \ \rm{erg\,s^{\rm{-1}}}$ and $L_{\rm{Q}} = \left(10^{41} - 10^{43}\right)\ \rm{erg\,s^{\rm{-1}}}$. Moreover, their central BH masses are inferred by the galaxy scaling relations \citep{Arcodia24b}.

Similar to QPEs, ASASSN-14ko, belonging to a subset called repeated optical nuclear transients (RNOTs), is a famous QPE-like event. It shows repeated optical flares with $P \approx 115\  \rm{days}$ in the nuclear of the galaxy \citep{payne21}. The flares last for $t_{\rm{e}} = \left(10 - 36\right)$ days with $L_{\rm{p}} \approx (0.3 - 10) \times 10^{44}\ \rm{erg\,s^{\rm{-1}}}$ \citep{payne21, payne22, payne23, huang23}. Spectral fitting gives $k T_{\rm{p}} = \left(0.3-2.5\right)\ \rm{eV}$ and the quiescent luminosity is $L_{\rm{Q}} = 2 \times 10^{42-43}\ \rm{erg\,s^{\rm{-1}}}$. Based on the relations between $M_{\rm{bulge}}-M_{\rm{BH}}$ and Two Micron All Sky Survey photometry, \cite{payne21} obtained $M_h \approx 7\times 10^7\ \rm{M}_{\rm{\odot}}$.

Furthermore, J0230 is another QPE-like source in the X-ray band with $P\ \approx\ 22\ \rm{days}$ \citep{evans23, guolo24}. Its period is much longer than that of QPE but shorter than that of RNOT. It peaks with $L_{\rm{p}}\ \approx\ 10^{42}\ \rm{erg\,s^{\rm{-1}}}$ and $t_{\rm{e}} = \left(2.4-6.5\right)\ \rm{days}$ and its $L_{\rm{Q}} = \left(1-2\right)\times 10^{41}\ \rm{erg\,s^{\rm{-1}}}$. Except for the peculiar period of flares, its $k T_{\rm{p}}\ =
\ 150-200\ \rm{eV}$ and the mass of BH inferred from the $M_{\rm{h}}-\sigma_*$ scaling relation is about $\left(1.5 \times 10^6 - 1\times 10^7\right)\ M_{\rm{\odot}}$ \citep{guolo24}.

In Table \ref{table:1}, we collect several observable properties ($P$, $t_{\rm{e}}$, $L_{\rm{p}}$, $L_{\rm{Q}}$ and $kT_{\rm{p}}$) for a sample of QPEs and QPE-like events and their host galaxies' central BH mass $M_{\rm{h}}$ estimated in the literature. For QPE sources that have been monitored for long-time coverage, the three most important observables ($t_{\rm{e}}$, $L_{\rm{p}}$ and $L_{\rm{Q}}$) usually vary from eruption to eruption, and therefore their observed values each have a range. In order to simplify our analysis in $\S$ \ref{sec:4}, we will use typical values for these observables.

\section{Constraints for the Model}
\label{sec:3}

Following each star-disk collision, the shocked material is brought out of the accretion disk, and then it will experience adiabatic expansion and cooling until the matter becomes mildly optically thin, after which the radiation trapped in the shocked matter can escape. If the star is moving in an orbit with a mild eccentricity, the star will collide with the disk twice in one orbital period. 

Based on the Kepler third law, assuming that the stellar orbit is near-circular, one can write disk radius $R$ at the collision point as

\beq 
R = \left[GM_{\rm{h}} \left(\frac{P_{\rm{orb}}}{2\pi} \right)^2 \right]^{\frac{1}{3}} = 110 P_{\rm{1}}^{\frac{2}{3}} M_{\rm{6}}^{-\frac{2}{3}}\ R_{\rm{g}},
\label{eq:r}
\eeq
where $R_{\rm{g}}=\frac{GM_{\rm{h}}}{c^2}$ is the gravitational radius of the BH and $c$ is the speed of light in vacuum. In order to simplify the derivation, we define the following notation: $r_\star = R_\star / R_{\rm{\odot}}$, $M_{\rm{6}} = M_{\rm{h}} / 10^6 \ M_{\rm{\odot}}$, $P_{\rm{1}} = P_{\rm{orb}} / 10\ \rm{hr}$ and $\alpha_{-1} = \alpha / 0.1$, where $R_{\rm{\odot}}$ and $M_{\rm{\odot}}$ are the solar radius and mass, $\alpha$ is the dimensionless disk viscosity parameter, and $R_\star$ is the radius of the star. 

\subsection{The Disk}

The disk middle plane speed of sound is $c_{\rm{s}} = \left( \frac{\partial P}{\partial \rho} \right)^{\frac{1}{2}} = \left(\gamma \frac{P}{\rho} \right)^{\frac{1}{2}}$, where $\gamma$ is the adiabatic index, $\rho$ is the density, and pressure $P = P_{\rm{gas}} + P_{\rm{rad}}$ is the combination of the gas pressure $P_{\rm{gas}}$ and the radiation pressure $P_{\rm{rad}}$. Here, we assume that the gas obeys the polytropic relation $P \propto \rho^{\gamma}$.

For the inner region of the disk where it is radiation-pressure and radiative-cooling dominated, the scale height of the disk is given by \citep[also see][]{frank2002,linial23}

\beq 
h = R\frac{c_{\rm{s}}}{v_{\rm{k}}} \simeq \frac{3 \kappa \dot M}{8 \pi c} \left[1 - \left( \frac{2 R_{\rm{g}}}{R}\right)^{\frac{1}{2}} \right] \approx 1.5  \dot m \ R_{\rm{g}},
\label{eq:h}
\eeq
where $v_{\rm{k}} = \sqrt{\frac{GM_{\rm{h}}}{R}}$ is the Kepler velocity at radius $R$, $\kappa$ is the opacity, and $\dot M$ is the mass accretion rate at the collision point. Hereafter, we consider that the opacity is dominated by Thompson scattering, thus $\kappa = 0.34\ \rm{cm}^2\ \rm{g}^{-1}$. The last step of the above has used $R\ \gg\ 2 R_{\rm{g}}$. We use the lowercase letter $\dot m \equiv \dot M / \left( L_{\rm{Edd}}/c^2 \right)$ to denote the dimensionless accretion rate,  where $L_{\rm{Edd}}$ is the Eddington luminosity. Then, the luminosity created by accretion is $L = \eta \dot m L_{\rm{Edd}}$, where $\eta$ is the radiation efficiency. 

Based on the $\alpha$-disk model \citep{ss73}, the surface density is given as

\beq 
\Sigma_{\rm{d}} = \frac{\dot M}{3\pi \nu} = 2 \times 10^{4} \frac{P_{\rm{1}}}{\alpha_{-1} \dot m M_{6}} \ \rm{g\,cm^{\rm{-2}}},
\label{eq: surface density}
\eeq
where $\nu \approx \alpha c_{\rm{s}} h \approx \alpha \sqrt{GMR} (\frac{h}{R})^2$ is the viscosity of the disk.

Then, the mass of shocked matter per collision is expressed as

\begin{align}
    M_{\rm{sh}}&=2\pi R_\star^2 \left( \sqrt{2} \Sigma_{\rm{d}} \right) \nonumber \\&= 4.3 \times 10^{-7} r_\star^2 \frac{P_{\rm{1}}}{\alpha_{-1} \dot m M_{6}} \ \rm{M}_{\rm{\odot}}.
\label{eq:msh}
\end{align}
The factor of $\sqrt{2}$ comes from the consideration that the velocity of the disk material is perpendicular to the stellar velocity and both are of the order of $v_{\rm{k}}$, so the effective length of the stellar trajectory is $2\sqrt{2} h$.

\subsection{Hydrodynamics in Collision}
\label{sec:hydrodynamics}

In the reference frame of the star, the relative velocity of the disk material is $\sqrt{2} v_{\rm{k}}$. Assuming all the kinetic energy of the unshocked matter is converted into the internal energy of the shocked one, the internal energy per unit mass of the shocked material is 

\beq
\epsilon_{\rm{sh}}/\rho_{\rm{sh}} = v_{\rm{k}}^2,
\label{eq:epsilon}
\eeq
where $\epsilon_{\rm{sh}}$ and $\rho_{\rm{sh}}$ are the internal energy density and the mass density of the shocked matter, respectively.

Taking into account $\gamma = \frac{4}{3}$ and the relation $P = \left(\gamma-1\right) \epsilon$, the speed of sound of the immediately postshock material is

\beq
c_{\rm{s,0}} = \sqrt{\gamma \frac{P_{\rm{sh}}}{\rho_{\rm{sh}}}} = \frac{2}{3} \ v_{\rm{k}},
\eeq
where $P_{\rm{sh}}$ is the pressure of the shocked matter. After a quasi-spherical adiabatic expansion, the terminal expansion speed of the entire shocked material would be $v_{\rm{ej}} \approx c_{\rm{s,0}}$.

\subsection{Duration of Emission}

By analogy to the supernova explosion, the duration of each eruption can be estimated by the time the optical depth of the shocked matter $\tau=\kappa \Sigma$ drops to the critical value of $c / v_{\rm{ej}}$ \citep{arnett80}. Then, most of the photons will escape from the shocked matter. This corresponds to when the flare reaches its peak luminosity. Hence, $\kappa \Sigma = c / v_{\rm{ej}}$, where

\beq 
\Sigma=\frac{M_{\rm{sh}}}{4 \pi R_{\rm{diff}}^2}
\label{eq:sigma_diff}
\eeq 
is the surface density of the fireball, $R_{\rm{diff}} = v_{\rm{ej}} t_{\rm{e}}$ is the radius of the fireball at its peak, and $t_{\rm{e}}$ is the duration of the eruption. Therefore, one has \citep[also see][]{linial23}

\beq 
t_{\rm{e}} \approx \left(\frac{\kappa M_{\rm{sh}}}{4\pi c v_{\rm{ej}}} \right)^{\frac{1}{2}} \approx 0.18 \frac{r_\star}{\alpha_{-1}^{\frac{1}{2}} \dot m^{\frac{1}{2}}}  \left(\frac{P_{\rm{1}}}{M_{\rm{6}}} \right)^{\frac{2}{3}}\  \rm{hr}.
\label{eq:tQPE}
\eeq

\subsection{Luminosity}

Next, we consider the energy evolution of the shocked matter. The following Eq. (\ref{eq:epsilon}), the total internal energy of the immediately postshock matter, which is dominated by radiation, is $E_{\rm{i,0}} = M_{\rm{sh}} v_{\rm{k}}^2$. Note that it is on the same order of magnitude as the total kinetic energy of the shocked matter $E_{\rm{k}} = \frac{1}{2} M_{\rm{sh}} v_{\rm{k}}^2$ \citep{linial23}.

During expansion, the temperature of the shocked matter decreases and a large fraction of the internal energy is converted into the expansion kinetic energy of the shocked matter. With $\gamma=\frac{4}{3}$, the internal energy evolves with the volume as $E_{i} \propto V^{1-\gamma}$.  So, the internal energy of the shocked matter at $R_{\rm{diff}}$ is 

\begin{align}
E_{i}\left( R_{\rm{diff}} \right) &= E_{\rm{i,0}}\times \left( \frac{2\pi R_\star^2 \sqrt{2} h/7}{\frac{4\pi}{3}R_{\rm{diff}}^3} \right)^{\frac{1}{3}} \nonumber\\
&= 3.9 \times 10^{44} r_\star^{\frac{5}{3}} \dot m ^{-\frac{1}{6}} \alpha_{\rm{-1}}^{-\frac{1}{2}} M_{\rm{6}}^{\frac{1}{3}} \ \rm{erg\,s^{\rm{-1}}}.
\end{align}
Note that the factor of $\frac{1}{7}$ in the numerator of the second equation comes from the shock compression during collision. Therefore, the peak luminosity of the diffusive radiation from the expanding shocked matter is

\beq
L_{\rm{p}}=\frac{E_{i}\left( R_{\rm{diff}} \right)}{t_{\rm{e}}} = 6.1 \times 10^{41}  M_{\rm{6}} \left( \frac{r_\star}{P_{\rm{1}}} \right)^{\frac{2}{3}} \dot m^{\frac{1}{3}}\ \rm{erg\,s^{\rm{-1}}}
\label{eq:Lp}.
\eeq
Note that it does not depend on $\Sigma_{\rm{d}}$, thus the parameter $\alpha$ vanishes.

\subsection{The Quiescent-level Accretion}

The quiescent-level X-ray flux between two neighboring eruptions can be considered to be emitted from the inner region of the accretion disk. Assuming the accretion rate during the quiescent state is not much different from that during the collision, one can write the quiescent-level luminosity as

\beq
L_{\rm{Q}} = 1.5 \times 10^{43} \eta_{-1} M_{6} \dot m\  \rm{erg\,s^{\rm{-1}}},
\label{eq:LQ}
\eeq
where $\eta = 0.1\times \eta_{-1}$ is the radiative efficiency.

\subsection{Infer $r_\star$ from Observables}

Eqs. (\ref{eq:tQPE}), (\ref{eq:Lp}), and (\ref{eq:LQ}) give the three observables $(t_{\rm{PQE}}$, $L_{\rm{p}}$ and $L_{\rm{Q}})$ from the shock-cooling emission model induced by star-disk collisions. Here, we attempt to investigate what $r_\star$ is allowed or required by the model to satisfy these observables.

Notice that $\alpha$ and $\eta$ are the two fudge ``factors" that describe the working mechanism of the accretion disk, and $\dot m$ is one of the disk properties; all others are observable or measurable in principle. One can get rid of $\dot m$ and express $r_\star$ in terms of observables in two ways. First, one can infer $\dot m$ from Eq. (\ref{eq:Lp}) and plug it into Eq. (\ref{eq:tQPE}) to get

\beq
r_\star= 4 L_{\rm{p,42}}^{\frac{3}{4}} P_{\rm{1}}^{\frac{1}{6}}M_{\rm{6}}^{-\frac{5}{12}} \left( \frac{t_{\rm{e}}}{1\ \rm{hr}} \right)^{\frac{1}{2}}{\alpha_{-1}}^{\frac{1}{4}}.
\label{eq: combine}
\eeq

Secondly, we can obtain $\dot m$ from $L_{\rm{Q}}$ (Eq. \ref{eq:LQ}) and plug it into Eq. (\ref{eq:Lp}), to get

\beq
r_\star = 32\ L_{\rm{p,42}}^{\frac{3}{2}} L_{\rm{Q,41}}^{-\frac{1}{2}} \frac{P_{\rm{1}}}{M_{\rm{6}}} \eta_{\rm{-1}}^{\frac{1}{2}},
\label{eq:lp-lq}
\eeq
where $L_{\rm{Q,41}} = L_{\rm{Q}}/\left( 10^{41} \  \rm{erg\,s^{\rm{-1}}}\right)$ and $L_{\rm{p,42}} = L_{\rm{p}}/\left( 10^{42} \right.$ $\left. \rm{erg\,s^{\rm{-1}}}\right)$.
In this way, we obtain two constraints on $r_\star$, both of which must be satisfied, and each contains one fudge factor only.

\subsection{Radiation Temperature}


One could also estimate the color temperature $T_{\rm{p}}$ of the radiation from the shocked material as another key observable. 

When the star-disk collisions occur, the material shocked by the collision will be radiation dominated and optically thick. 
Firstly, the radiation energy density is expressed as $u_{\gamma} = L_{\rm{p}} / [4\pi R_{\rm{diff}}^2 v_{\rm{ej}}]$ \citep{linial23}. Secondly, assuming that the radiation of QPEs is dominated by blackbody radiation, $u_{\gamma} = a T_{\rm{p}}^4$, where $a$ is the radiation constant, then $T_{\rm{p}}$ can be expressed in terms of various observables as \citep[also see][]{linial23}

\begin{align}
    k T_{\rm{p}} = 9 \ L_{\rm{p,42}}^{\frac{1}{4}} \left(\frac{t_{\rm{e}}}{1\ \rm{hr}}\right)^{-\frac{1}{2}} P_{1}^{\frac{1}{4}} M_{\rm{6}}^{-\frac{1}{4}} \ \rm{eV},
\label{eq:Tp}
\end{align}
where $k$ is the Boltzmann constant.

Note that \cite{linial23} argue that if the radiation is not in thermal equilibrium with the matter at $R_{\rm{diff}}$, then $k T_{\rm{p}}$ might be higher.

\subsection{Constraint from Stellar Tidal Disruption}

Now we consider another constraint on $r_\star$. The stellar orbit must be sufficiently wide; otherwise the star would be tidally disrupted. A critical orbital radius is the so-called tidal disruption radius \citep[e.g.,][]{rees88}:
\beq
R_{\rm{t}} = R_\star \left(\frac{M_{\rm{h}}}{M_*}\right)^{\frac{1}{3}} = 47 r_\star M_{\rm{6}}^{-\frac{2}{3}} m_*^{-\frac{1}{3}}\ R_{\rm{g}}.
\label{eq:rt}
\eeq
The above constraint requires that $R>nR_{\rm{t}}$, where $n=1$ corresponds to the full disruption and $n=2$ corresponds to the starting point of mass transfer; in other words, $1<n<2$ corresponds to the partial disruption of the star.


Plugging Eq. (\ref{eq:r}) in, this can be rearranged to put a constraint on the stellar radius: $r_\star < 2.4\ n^{-1} P_{\rm{1}}^{\frac{2}{3}} m_*^{\frac{1}{3}}$. For a normal star, one can adopt the mass-radius relationship for high-mass zero-age-main-sequence stars : $r_\star = m_*^{0.6}$ \citep{kw13}. However, as was mentioned in \cite{linial23}, \cite{linial23a}, and \cite{linial24} and was shown in simulation by \cite{yao25}, the star may get heated and inflated during collisions. The increased $r_{\star}$ would cause $R_{\rm{t}}$ in Eq. (\ref{eq:rt}) to be larger,  making it easier to be disrupted. Then the condition of $R > nR_{\rm{t}}$ would exclude a boarder parameter space.

To include this effect, we introduce a bloating factor $b$ into the mass-radius relations for an inflated star: $r_\star = b m_*^{0.6}$. Then, one can rewrite the condition for avoiding the disruption as

\beq
r_{\rm{*}} < 6.9\ n^{-\frac{9}{4}} P_{1}^{\frac{3}{2}} b^{-\frac{5}{4}}.
\label{eq:r-star-max}
\eeq






 \begin{figure*}[h]
   \centering
    \includegraphics[width=3in]{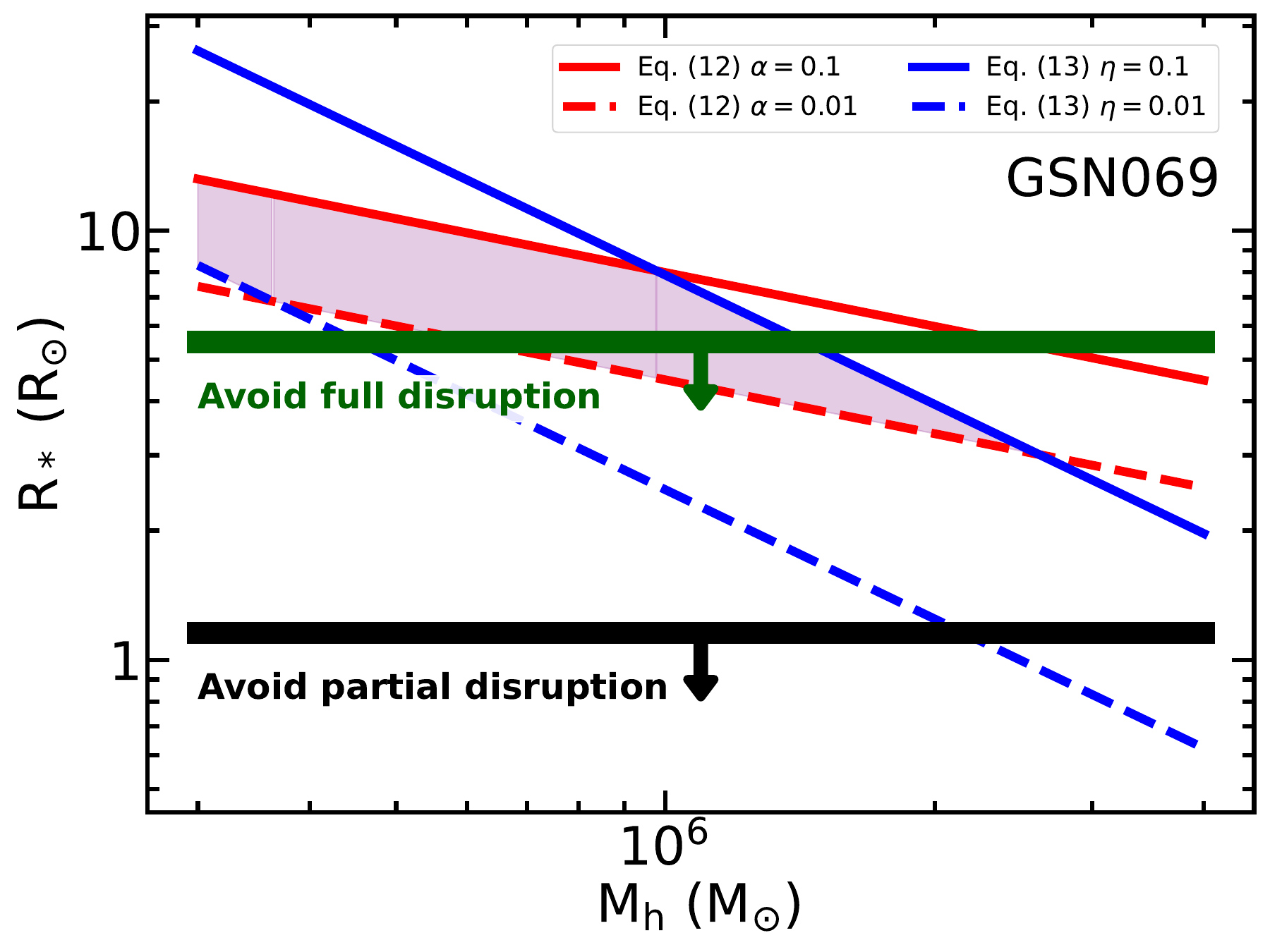} 
    \includegraphics[width=3in]{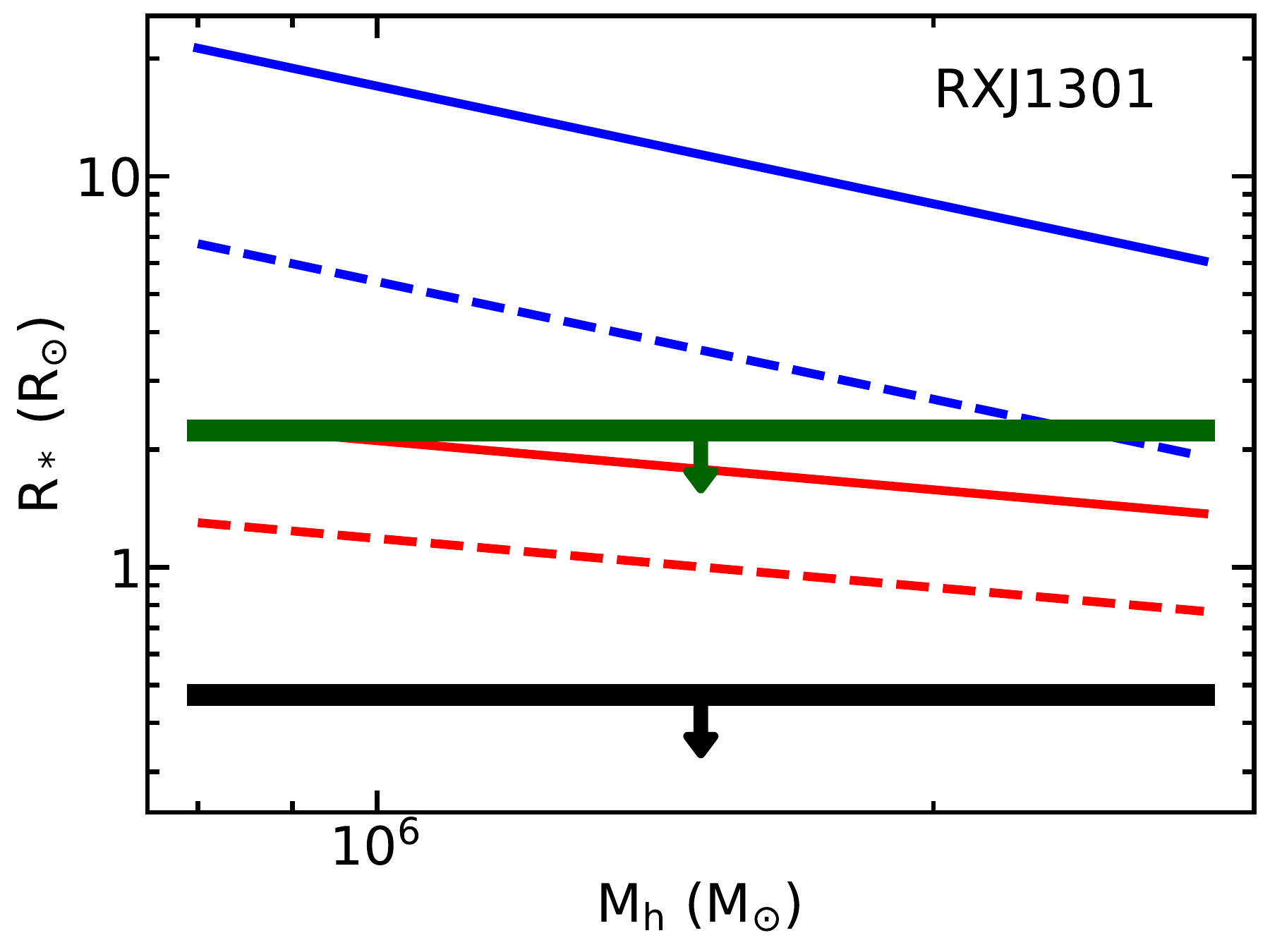}    
   \caption{The constraints on the stellar radius for GSN 069 (left panel) and RX J1301 (right panel). The red and blue lines stand for the constraints based on Eq. (\ref{eq: combine}) and Eq. (\ref{eq:lp-lq}), respectively. The solid and dashed lines are for different adopted values of $0.1$ and $0.01$ for the model parameter ($\alpha$ or $\eta$), respectively. The horizontal lines with downward arrows represent the full tidal disruption limit (green, Eq. \ref{eq:r-star-max}, $n = 1$ and $b=1$) and the partial tidal disruption limit (black, Eq. \ref{eq:r-star-max}, $n=2$ and $b=1$), respectively.}
   \label{fig:gsn}
\end{figure*}

\begin{figure*}[h]
    \centering
    \includegraphics[width=6.2in]{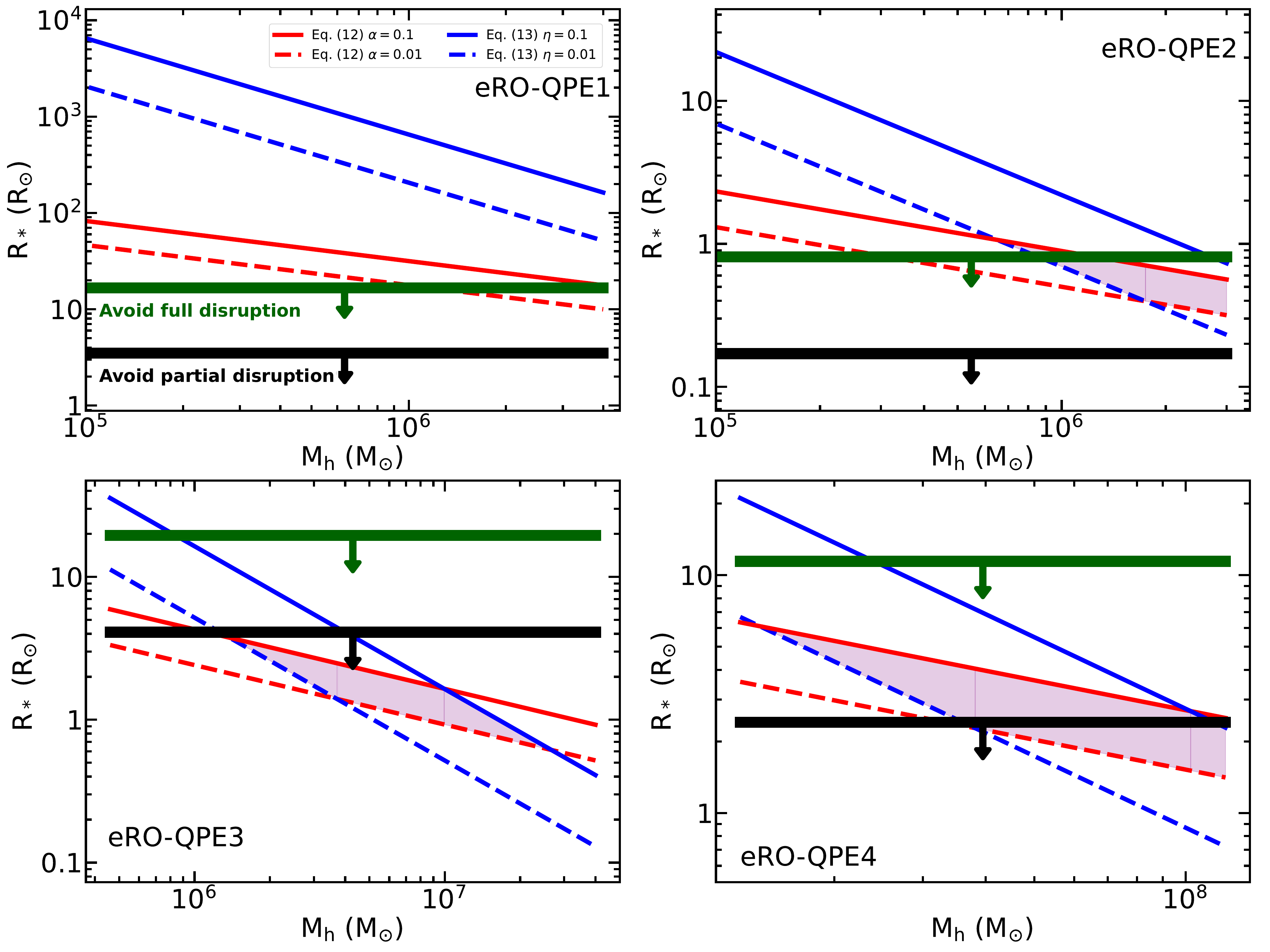}  
   \caption{The constraints on the stellar radius for other four QPE sources. The legends are the same as those in the left panel of Figure \ref{fig:gsn}.}
   \label{fig:other-qpe}
\end{figure*}

 \begin{figure*}[htbp]
   \centering

    
   \includegraphics[width=3.1in]{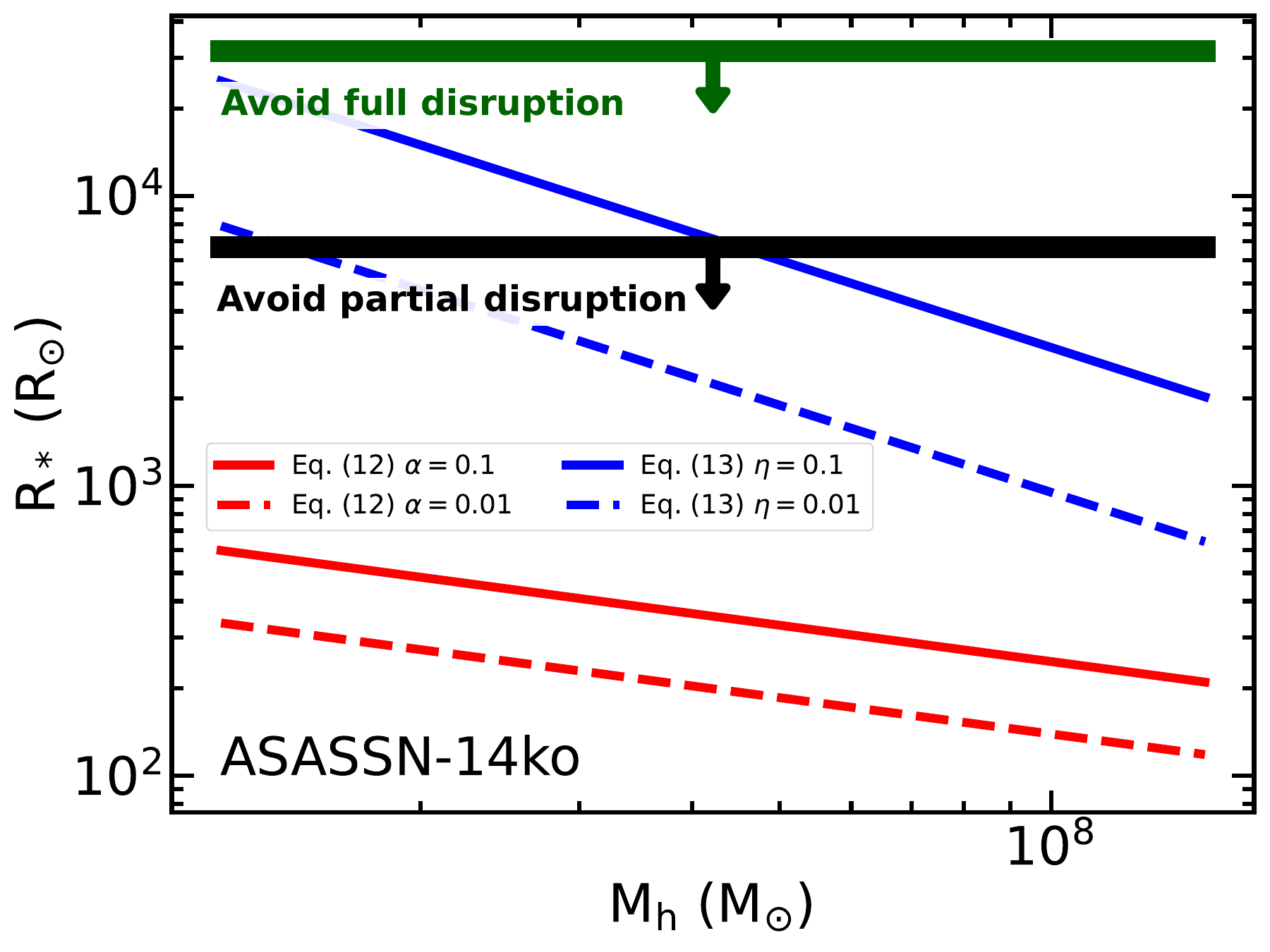}       
   \includegraphics[width=3.1in]{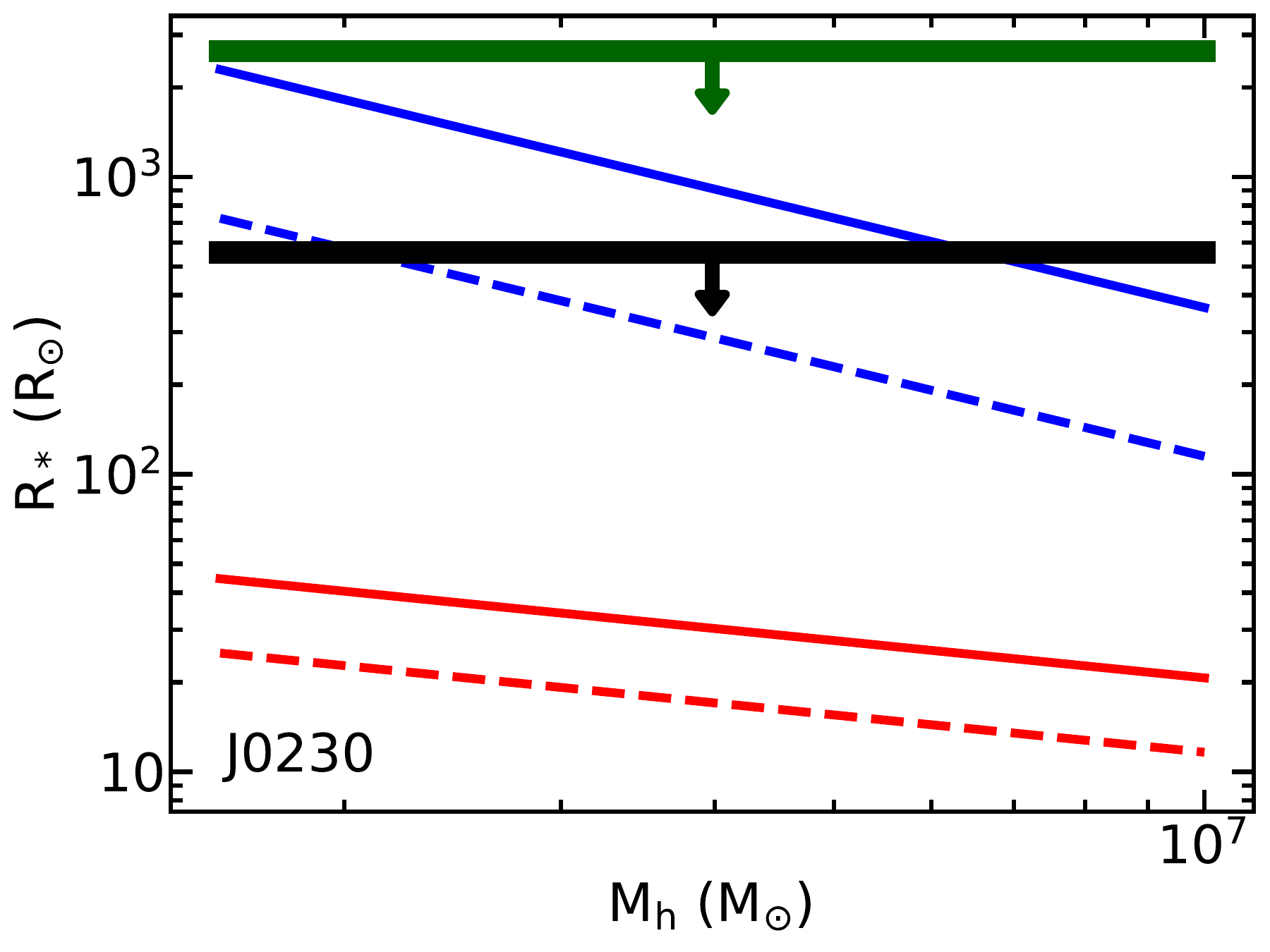} 
   \caption{The constraints of stellar radius for longer time-scale repeating sources: ASASSN-14ko and J0230. The legends are the same as those in the left panel of Figure \ref{fig:gsn}.}
   \label{fig:other-src}
\end{figure*}

\section{Application to QPE Sample}
\label{sec:4}

In this section, for each source, we plot the $r_\star$-constraints as functions of $M_{\rm{h}}$. The numerical range of $M_{\rm{h}}$ of an individual source corresponds to the estimate quoted in Sect. \ref{sec:2} and Table. \ref{table:1}.

\subsection{GSN 069}

We start with the well-known QPE source GSN 069. In Figure \ref{fig:gsn}, we plot the $r_\star$-constraints, Eq. (\ref{eq: combine}) and Eq. (\ref{eq:lp-lq}) as the red and blue lines, respectively. The solid and dashed lines are for two adopted representative values of the unknown parameters ($\alpha$ and $\eta$), respectively. The light-purple shadowed region represents the allowed parameter space that could satisfy both Eq. (\ref{eq: combine}) and Eq. (\ref{eq:lp-lq}). The full tidal disruption limit and the partial tidal disruption limit are represented by the green and black horizontal lines, respectively. The region below the green line means that stars with such small $r_\star$ can avoid being tidally disrupted. Similarly, stars with even smaller $r_\star$ (below the black line) will not even be tidally stripped.

As the left panel of Figure \ref{fig:gsn} shows, GSN 069 does have a light-purple shaded region of $r_\star \lesssim 10$, especially for $M_{\rm{h}} \lesssim 10^6\ M_{\rm{\odot}}$, that satisfies both constraints. However, most of that region is above the solid green line, meaning that they would have already been tidally disrupted. 

Only a small shaded region with small $r_\star$ (i.e., below the green solid line) can avoid the full disruption, but they are still well above the black solid line, meaning that they would suffer a significant partial disruption and that disruption would repeat on every orbit. In the latter case, the question arises that whether the stripped mass and its subsequent accretion toward the BH will be the dominant energy source for the observed eruption. Indeed, a few groups have argued for the repeated strippings or mass transfer from a stellar companion as the origin of QPEs \citep{king20, king22, king23, zhao22, chen23, wu24, krolik22, lu23}. These results demonstrate that the simplest version of the shock-cooling emission model from star-disk collisions did not pass our test for GSN 069.



\subsection{Other QPE Sources}

Next, we present results for other QPE sources. The constraints for RX J1301 are shown in the right panel of Figure \ref{fig:gsn}. It shows that this source does not even have an allowed $r_\star$-parameter space that can satisfy both Eq. (\ref{eq: combine}) and Eq. (\ref{eq:lp-lq}). Considering an extremely low $\eta$ (e.g., $\sim 10^{-3}$) might allow a light-purple shaded region to appear, but it will still be above the black solid line, meaning that the star would suffer repeated partial disruptions. Therefore, the shock-cooling emission model from star-disk collisions can work for RX J1301.

Figure \ref{fig:other-qpe} shows the constraints for other four QPE sources. One can see that for eRO-QPE1, there is no parameter space that satisfies both constraints Eq. (\ref{eq: combine}) and Eq. (\ref{eq:lp-lq}). For eRO-QPE2, a small parameter space of $r_\star \sim 0.5$ for $M_{\rm{h}} \gtrsim 10^6\ M_{\rm{\odot}}$ is allowed by both Eq. (\ref{eq: combine}) and Eq. (\ref{eq:lp-lq}), but similar to the situation for GSN 069, it is well above the solid black line, so it will suffer significant partial tidal disruptions. 

eRO-QPE3 and eRO-QPE4 fare much better in our test, as shown in the lower panels of Figure \ref{fig:other-qpe}. The best case is eRO-QPE3, whose allowed parameter space is $r_\star \sim 1$ (i.e., a sun-like star) and $M_{\rm{h}} \lesssim 10^7\ M_{\rm{\odot}}$, and it is free from any tidal stripping. eRO-QPE4 also has an allowed parameter space, and taking into account the tidal stripping limit, that space shrinks to $r_\star \sim 2$ and $M_{\rm{h}} \sim 10^8 \ M_{\rm{\odot}}$.

Assuming an inflated star whose bloating factor is $3$ \citep{yao25}, Eq. (\ref{eq:r-star-max}) becomes $r_{\rm{*}} < 1.3\ n^{-\frac{9}{4}} P_{1}^{\frac{3}{2}} \left(\frac{b}{3}\right)^{-\frac{5}{4}}$. Compared with the case of no inflation $\left(b=1\right)$, its full or partial tidal disruption limit would provide a more squeezed parameter space because an inflated star would be easier to get disrupted.

\subsection{QPE-like Sources with Longer Periods}  \label{sec:othersource}

Now we turn to those QPE-like sources with relatively long periods: ASASSN-14ko and J0230, whose results are shown in Figure \ref{fig:other-src}. Firstly, because these two sources have much longer time scales $t_{\rm{e}}$ and $P$, they require extremely large $r_\star$ to power the luminous flares, as can be seen in Eqs. (\ref{eq:tQPE}) and (\ref{eq:Lp}). In some cases, the required $r_\star$ (i.e., $10^{3-4}$ by Eq. \ref{eq:lp-lq}) is too large to be physically reasonable. Secondly, although a large $P$ -- hence a wide orbit -- puts such a star safe from a full or partial tidal disruption, there is hardly a common parameter space that can meet both Eqs. (\ref{eq: combine}) and (\ref{eq:lp-lq}). We note that a repeated partial tidal disruption model has been invoked to explain ASASSN-14ko \citep{payne21, payne22, payne23, huang23}.


\subsection{Comparison of $\ T_{\rm{p}}$}

 \begin{figure}[ht]
   \centering
   \includegraphics[width=3in]{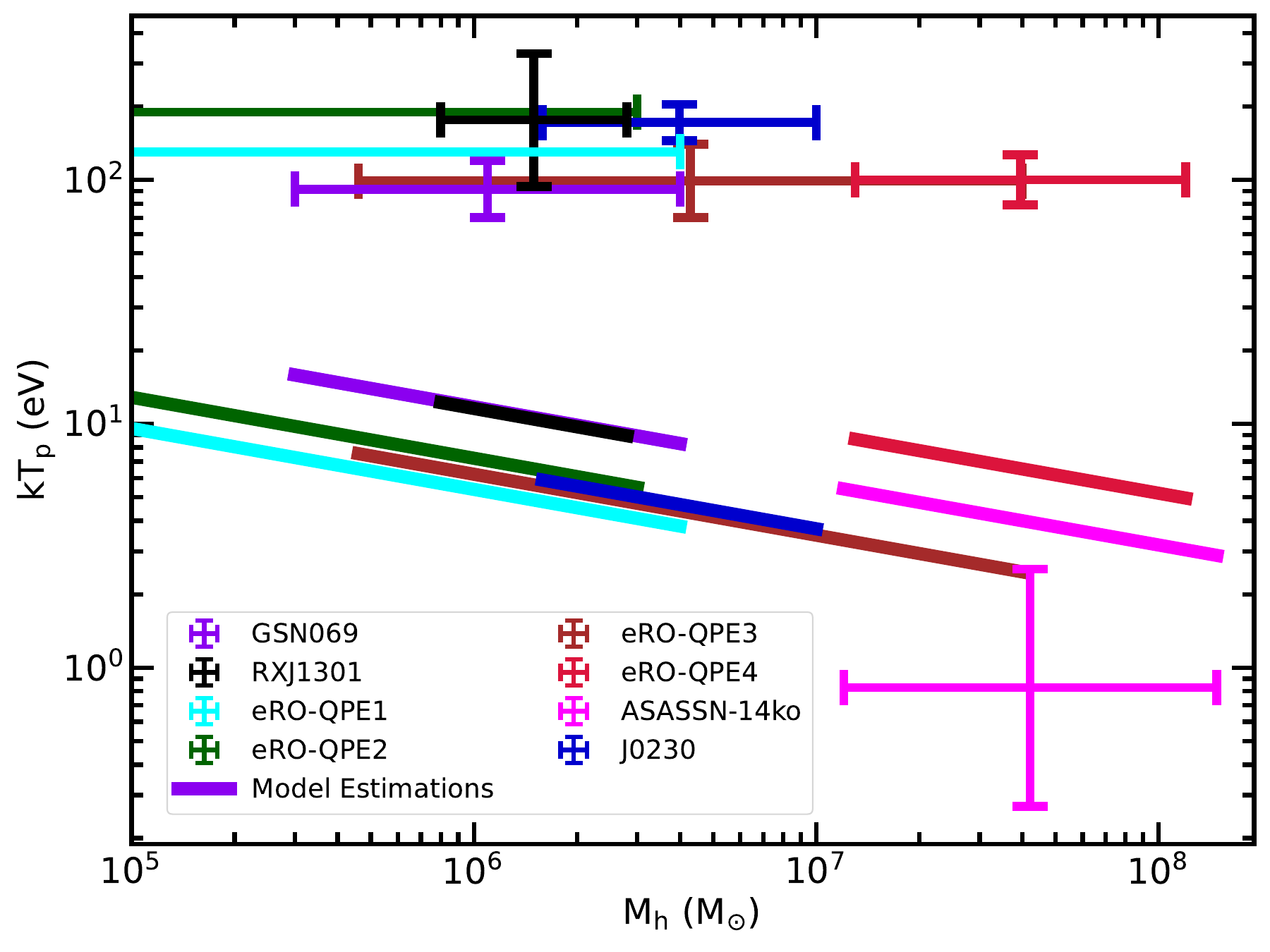} 
   \caption{The comparison of the peak radiation temperatures $T_{\rm{p}}$ predicted by the shock-cooling emission model from star-disk collisions (solid lines) and those observed (crosses with error bars) for the sample. The predicted $T_{\rm{p}}$ is calculated from Eq. (\ref{eq:Tp}). The horizontal error bars of the observed $T_{\rm{p}}$ correspond to the estimated $M_{\rm{h}}$ range of each individual source.}
   
   \label{fig: tp}
\end{figure}

We also compare the predicted peak radiation temperature $T_{\rm{p}}$ (Eq. \ref{eq:Tp}) with the observed ones, which are shown in Figure \ref{fig: tp}. It is obvious that, except for ASASSN-14ko, the predicted $T_{\rm{p}}$'s for all other sources are 1 order of magnitude lower than the observed values.



Note that Eq. (\ref{eq:Tp}) is based on the assumption that the radiation is fully thermalized with the matter when it is diffusing out of the expanding shocked material. \cite{linial23} argue that it may not be the case, so that $T_{\rm{p}}$ could be higher.


For ASASSN-14ko, the observed $k T_{\rm{p}} \sim 1 \ \rm{eV}$ is low, which is not surprising since its radiation during each eruption is dominated by the optical wave band \citep{payne21, payne22, payne23, huang23}. Thus, the discrepancy does not appear to be an issue. However, we have shown in \S \ref{sec:othersource} that the shock-cooling emission model from star-disk collisions cannot pass the $r_\star$ test for this source, thus the model can be ruled out.


\section{Discussion}
\label{sec:5}

\subsection{Orbital Inclination Effect}
\label{sec:inclination}

Our above analysis so far has assumed a vertical stellar orbit, i.e., the inclination angle $i$ between the angular momenta of the star and the disk is $\frac{\pi}{2}$. Here, we consider a general case whose $i$ has other values.



In the reference frame of the star, the relative velocity of the disk is $\sqrt{2\left(1 -\cos{i}\right)} v_{\rm{k}}$ and the effective length of the stellar in-disk trajectory is $\min \left[ 2\sqrt{\frac{2}{1 +\cos{i}}}h, \ 2 \pi R \right]$. 
For an extremely prograde orbit ($i \approx 0$), the collisions would be very weak. An intermediate inclination ($\frac{\pi}{4} \lesssim i \lesssim \frac{3\pi}{4}$) would only introduce an order-of-unity change to our previous results. Thus, for the demonstrative purpose, here we consider an extremely retrograde orbit ($i \approx \pi$) as the most optimistic case.

In this case, the relative velocity is $2 v_{\rm{k}}$, the effective length of the stellar trajectory is $\approx 2\pi R$ and the cross section is $\pi R_\star \times \min \left[R_\star, h\right]$. 
From Eq. (\ref{eq:h}), one has $h \approx 3 \dot m M_{\rm{6}}\ R_{\rm{\odot}}$. Considering two regimes, $h > R_\star$ and $h < R_\star$, one could derive the duration and the peak luminosity, similarly to in $\S$ 3.3 - 3.4, as

\begin{equation}
    t_{\rm{e}} = 
        \left\{
            \begin{array}{@{}l@{\quad}l@{}}
               1.9\ r_\star P_{\rm{1}} M_{\rm{6}}^{-1}\alpha_{\rm{-1}}^{-\frac{1}{2}}\dot m^{-1} \ \rm{hr}, & \text{for } h > R_\star, \\[3ex]
               3.4\ r_\star^{\frac{1}{2}} P_{\rm{1}}  M_{\rm{6}}^{-\frac{1}{2}} \alpha_{-1}^{-\frac{1}{2}} \dot m^{-\frac{1}{2}} \ \rm{hr}, & \text{for } h < R_\star,
            \end{array}
        \right.
        \label{eq:te-lowinclination}
\end{equation}
and
\begin{equation}
L_{\rm{p}} = 
    \left\{
        \begin{array}{@{}l@{\quad}l@{}}
            6.8 \times 10^{42}\ r_\star^{\frac{2}{3}} P_{\rm{1}}^{\frac{2}{9}} M_{\rm{6}}^{\frac{1}{9}}\ \rm{erg\,s^{\rm{-1}}}, & \text{for } h\ >\ R_\star, \\[3ex]
            1 \times 10^{43}\ r_\star^{\frac{1}{3}} P_{\rm{1}}^{\frac{2}{9}} M_{\rm{6}}^{\frac{7}{9}} \dot m^{-\frac{1}{3}}\ \rm{erg\,s^{\rm{-1}}}, &\text{for } h\ <\ R_\star.
        \end{array}
    \right.
    \label{eq:lp-lowinclination}
\end{equation}

Based on  Eqs. (\ref{eq:te-lowinclination}-\ref{eq:lp-lowinclination}) and using Eq.(\ref{eq:LQ}), one can derive two solutions of $r_\star$ in terms of observables, similarly to Eqs. (\ref{eq: combine}-\ref{eq:lp-lq}). One then checks whether the two solutions have an overlapping region, considering a numerical range of $\left(0.01, 0.1\right)$ for both $\alpha$ and $\eta$, and whether they satisfy the tidal constraint of Eq. (\ref{eq:r-star-max}), similarly to Figures \ref{fig:gsn}-\ref{fig:other-src}.

We found that three sources (GSN 069, eRO-QPE2, eRO-QPE3) have allowed $r_\star \sim 0.1$ in the $h > R_\star$ regime; one source (eRO-QPE1) has allowed $r_\star \sim 0.01$ in the $h < R_\star$ regime. The rest in our sample do not have any allowed $r_\star$.

Therefore, the extremely retrograde orbit is indeed a feasible solution to four sources. However, one should note that for a random orientation of the stellar orbit, the differential probability distribution of $i$ is $\propto \sin i$, such that the probability for $i \in \left(\frac{3\pi}{4}, \pi\right)$ is $\approx 15 \%$ \citep[also see][]{tomoya25}. Thus, the chances for this most optimistic case are low.



\subsection{Stream-Disk Collisions}

Here, we briefly consider the stream-disk collision, a derivative version of the original star-disk collision model, where the stream was shed from the star via ablation \citep{lu23, linial24, yao25, linial25} or tidal stripping \citep{lu23, huang23}.


Regarding the cross section of the stream, according to \cite{yao25}, its radial width is determined by the dispersion in orbital energy to be $\Delta R \approx 10\ P_{\rm{1}} m_\star^{\frac{1}{2}} r_\star^{-\frac{1}{2}}\ R_\odot \gg R_\star$, whereas its transverse size is $\approx R_\star$ (in \citealt{linial25}, numerically calculating the Hills equations yields an even more extended stream size). Therefore, when the stream crosses the disk, it has a much larger shocking area $\pi R_\star \Delta R$ than the star alone.

As long as the stream's column density is larger than the disk surface density, $\Sigma_{\rm{s}} = \frac{\Delta M_\star}{\pi R_\star \Delta R} \gtrsim \Sigma_{\rm{d}}$ \citep[this requires $P_{\rm{1}} \lesssim 6$, while other parameters are fixed at their fiducial values; see Eq. 18 of][]{yao25} where $\Delta M_\star$ is the stream's mass, 
 the stream-disk collision would shock more disk material and then power a brighter and longer-lived flare than the star-disk collision.

In this scenario, the framework in $\S$ \ref{sec:3} for constraining the shock-cooling emission still applies. To get the constraints on the stream's cross section, we can just substitute all term of $R_\star^2$ in equations there with $R_\star \Delta R$ and drop off the tidal disruption constraint Eq. (\ref{eq:r-star-max}). In other words, one can interpret $r_\star$ in Eqs. (\ref{eq: combine}-\ref{eq:lp-lq}) as the square root of this cross section in unit of $R_\odot$.

After removing the green and black horizontal lines in Figures \ref{fig:gsn}-\ref{fig:other-src}, one can see that four sources within our sample (i.e., GSN 069, eRO-QPE2, eRO-QPE3, and eRO-QPE4) have the allowed cross-section size that can simultaneously reproduce the observed $L_{\rm{p}}$, $t_{\rm{e}}$ and $L_{\rm{Q}}$. The rest does not pass our tests on radiative properties. We conclude that, under the condition of $\Sigma_{\rm{s}} \gtrsim \Sigma_{\rm{d}}$, the stream-disk collision can work for the majority of the QPE sources (excluding the two long-period QPE-like events).

On the other hand, if $\Sigma_{\rm{s}} < \Sigma_{\rm{d}}$ (i.e., $P_{\rm{1}} > 6$), the reverse shock propagating into the debris stream might be strong enough, and the shocked stream may dominate the emission and power the eruption. In this case, the extended radial size of the stream leads to a significant dispersion in its orbital energy. Thus, different segments of the stream arrive the disk plane at markedly different times. The spread of arrival times, $\Delta t_{\rm{s}} = \left[P_{\rm{orb}}\left(R+\Delta R\right) - P_{\rm{orb}}\left(R-\Delta R\right)\right]/2 \approx \frac{3}{2} \frac{\Delta R}{R} P_{\rm{orb}} \approx 6.5 m_\star^{\frac{1}{2}} M_{\rm{6}}^{-\frac{5}{6}} \left(\frac{P_{\rm{1}}}{6}\right)^{\frac{4}{3}}\ \rm{hr}$, would represent the effective collision time scale, which might be comparable or even larger than the expansion time scale (see Eq. \ref{eq:tQPE}) of the shocked stream material.

This indicates that, for the shocked stream case, the stream-disk collision is a prolonged collision, rather different from the almost instantaneous impact in the star-disk collisions. Hence, our framework presented in $\S$ \ref{sec:3} and the findings in $\S$ \ref{sec:4} would not apply for this case.

\subsection{Breakout Emission}

Besides the shock-cooling emission model, in analogy with a supernova explosion, the star-disk collision may give rise to a breakout emission \citep{tagawa23}. The breakout emission releases the internal energy on a short time scale without the adiabatic energy loss, which is expected to produce a more luminous flare. However, the duration of breakout emission is approximated by its dynamical time scale ($\sim h / v_{\rm{k}}$), which poses a challenge in producing a QPE flare with a typical duration of $1\ \rm{hr}$. Although \cite{tagawa23} addressed the duration issue by introducing a bow shock profile, more careful checks are required in future work.

\subsection{BH as the Collider}

Lastly, we briefly consider a compact object, such as a BH, which is free of the tidal disruption constraint, as the object that collides with the disk. 
In this scenario, the collider will tidally capture the disk material during each crossing. The capture radius is $R_{\rm{c}} = 2 G M_{\rm{*}}/v_{\rm{k}}^2$  $\approx 5 \times 10^{-4} m_* P_{\rm{1}}^{\frac{2}{3}} M_{\rm{6}}^{-\frac{2}{3}} \ R_{\rm{\odot}}$ and the captured mass is $M_{\rm{c}} = 2 \pi R_c^2 (\sqrt{2} \Sigma_d)$.



The most efficient way to release gravitational energy is the BH accretion. Assuming the captured material is completely accreted by the collider, the total energy released is $E_{\rm acc} = \eta M_{\rm{c}} c^2 \approx 1.5 \times 10^{40} \eta_{-1} \alpha_{-1}^{-1} \dot m^{-1} m_*^2 P_{\rm{1}}^{\frac{7}{3}} M_{\rm{6}}^{-\frac{7}{3}} \ \rm{erg}$.

The typically observed total radiation energy during one eruption from a QPE source is $E_{\rm{rad}} \simeq 7 \times 10^{45}\ \rm{erg}$ \citep{miniutti23a}. This would require $m_* \sim 10^3$, i.e., an intermediate-mass BH (IMBH), as the collider.
Alternatively, if a QPE is powered by a nonaccretion process, such as the shock heating and expansion cooling of the captured material, the effective radiative efficiency $\eta$ would be lower, thus the required $m_\star$ would be even higher.

The gravitational wave radiation driven orbital decay time scale of this IMBH-SMBH binary would be $\tau_{\rm{GW}} \approx 2\times 10^3 \left(m_*/10^3 \right)^{-1} M_{\rm{6}}^{-\frac{2}{3}} P_{\rm{1}}^{\frac{8}{3}}\ \rm{yr}$ \citep{shapiro1986,peter1963}.
 Although this is not inconsistent with any observational fact of QPEs, the rarity of direct observational evidence of IMBHs in general so far puts this possible scenario into question. More support and independent evidence are needed for this scenario.

\section{Conclusion}
\label{sec:6}

The star-disk collision model proposed for QPE and QPE-like sources has attracted a lot of attention and remains a hot topic of discussion. However, the matching between the observed radiative properties, including $L_{\rm{p}}$, $t_{\rm{e}}$ and $T_{\rm{p}}$, and the model predictions has not been fully examined. In our study, we derive constraints on $r_\star$ and $T_{\rm{p}}$ as functions of $M_{\rm{h}}$ for the shock-cooling emission model from star-disk collisions, using the typical values of these observables. We then apply this method to a sample of QPE and QPE-like sources, with a particular attention to what value or range of the stellar radius $r_\star$ is required to reproduce these observables. The following are our findings.



1) Within our sample, the eruption luminosity and duration constraints require that $r_\star \sim 3 - 10$ for GSN 069, $r_\star \sim 0.3 - 1$ for eRO-QPE2, $r_\star \sim 0.5 - 4$ for eRO-QPE3, and $r_\star \sim 2 - 5$ for eRO-QPE4. There is no allowed parameter space for the rest (RX J1301, eRO-QPE1, ASASSN-14ko, and J0230) of the sample.


2) Introducing the full or partial tidal disruption limit further squeezes the allowed parameter space of eRO-QPE4 to $r_\star \sim 2$. The best candidate source eRO-QPE3 satisfies this limit well. For GSN 069 and eRO-QPE2, however, they would have suffered significant partial disruption, and the accretion of the stripped mass might be the dominant energy source powering the eruptions, thus undermining the foundation of the shock-cooling emission model from star-disk collisions.

3) The influence of the orbital inclination $i$ is the most notable when the stellar orbit is extremely retrograde. In that optimistic case, a star with a small $r_\star$ of $0.01-0.1$ would be able to reproduce the observed $t_{\rm{e}}$ and $L_{\rm{p}}$ for four sources (i.e., GSN 069, eRO-QPE1, eRO-QPE2 and eRO-QPE3). However, it is rare for the system to have $i \sim \pi$.




4) There exists a systematic discrepancy between the model-predicted $T_{\rm{p}}$ $\sim \left(0.01\ \rm{keV}\right)$ and observed values $\sim \left(0.1\ \rm{keV}\right)$, though the incomplete thermalization of the diffusive radiation might be a remedy.

Our findings pose serious challenges to the shock-cooling emission model from star-disk collisions. Specifically, we find that for most of the sample, the requirements of $r_\star$ to reproduce both $L_p$ and $t_e$, and the requirement that the star should not be partially disrupted unless including the inclination effect, are in tension with each other. This issue is largely alleviated when the star is accompanied by a debris stream shed from the star, and the stream-disk collisions produce the eruptions.

We thank an anonymous reviewer for the useful comments and suggestions that helped improve the manuscript, and are grateful to the organizers and speakers of the TDE Full-process Orbital to Radiative Unified Modeling for the insightful discussions. This work is supported by the National Natural Science Foundation of China grant 12393814.


\bibliography{A64544v6}{}
\bibliographystyle{aasjournal}

\end{CJK*}
\end{document}